\documentclass[RNAAS]{aastex62}

\graphicspath{{./}{figures/}}

\begin{document}

\title{Spectral curvature of shock-accelerated particles in solar cycle $23$}

\correspondingauthor{Federico Fraschetti}
\email{federico.fraschetti@cfa.harvard.edu}

\author{Connie Zhou}
\affiliation{Harvard/Smithsonian Center for Astrophysics, 60 Garden St., Cambridge, MA, USA, 02138; Winsor School, Boston, MA, USA, 02215}

\author{Federico Fraschetti}
\affiliation{Harvard/Smithsonian Center for Astrophysics, 60 Garden St., Cambridge, MA, USA, 02138; Depts. of Planetary Sciences and Astronomy, Tucson, AZ, USA, 85721}

\author{Jeremy J. Drake}
\affiliation{Harvard/Smithsonian Center for Astrophysics, 60 Garden St., Cambridge, MA, USA, 02138}

\author{Martin Pohl}
\affiliation{DESY, 15738 Zeuthen, Germany; Institute of Physics and Astronomy, University of Potsdam, 14476 Potsdam, Germany}



\keywords{editorials, notices --- 
miscellaneous --- catalogs --- surveys}

\section{}

In the late '70s, the acceleration of charged particles at astrophysical shock waves was theoretically predicted, within the diffusive shock acceleration model, to result in a power-law momentum distribution of the accelerated particles, i.e., $N(\gamma) = N_0 (\gamma/\gamma_0)^{-\alpha}$, where $\gamma$ is the electrons Lorentz factor and $\alpha$ the power law index \citep{Bell:78a, Blandford.Ostriker:78}. Recently, \citet{Fraschetti.Pohl:17a} modelled the baseline photon spectrum of the Crab Nebula, the remnant of a supernova explosion that occurred in AD $1054$ and showed that a single log-parabola spectrum of energetic electrons, accelerated at the termination shock of the Nebula, can reproduce the observations over $21$ orders of magnitude in photon energy. The log-parabola takes the form $N(\gamma) = N_0 (\gamma/\gamma_0)^{-\alpha - \beta {\rm log}(\gamma/\gamma_0)}$, where $\beta$ is the spectral curvature. 

Double power-laws in solar energetic particle (hereafter SEPs) spectra have been observed for more than a decade \citep[e.g.][]{Mason.etal:02,Mewaldt.etal:05}. \citet{Li.Lee:15} have shown that the Parker equation with power-law injected particles near the Sun admits a double power-law as a solution at $1$~AU if one includes an energy dependence of the scattering mean free path and adiabatic deceleration in the radially divergent solar wind. However, that study neglects perpendicular diffusion \citep[e.g.][]{Fraschetti.Jokipii:11} that plays a relevant role in large gradual SEP events \citep[and references therein]{Droege.etal:16}. \citet{Mewaldt.etal:12} reported the proton spectra of the 16 Ground Level Events (hereafter GLEs), i.e., SEPs events with a proton flux exceeding the background of galactic cosmic rays in the neutron monitors \citep{Lopate:06}, that occurred during solar cycle $23$. The spectra of such possibly shock-accelerated protons (in the range $\sim 0.1$ MeV to $500-700$ MeV) can be reproduced \citep{Mewaldt.etal:12} by the Band function \citep{Band.etal:93}.  This is an empirical fitting function originally introduced to fit the $\gamma$-ray spectra of Gamma Ray Bursts, and comprises two power-laws with different index continuously joined at an energy break $E_0$. 

In this study, we have used the GLE spectra of solar cycle $23$ to establish a relation of  $\beta$ to $E_0$. We assume that the measured spectra do not change since the injection, likely to happen in shocks driven by gradual coronal mass ejections. The log-parabola spectrum results from the leakage of the highest-energy particles that leads to the spectral softening and is described in terms of the probability of containing such particles in proximity of the shock \citep{Massaro.etal:04,Fraschetti.Pohl:17a, Fraschetti.Pohl:17b}. As the particle energy grows, the non-planarity of the shock surface and the greater scattering mean free path modify this probability by introducing an energy-dependence. 

\begin{figure}[h!]
\begin{center}
\includegraphics[scale=0.7,angle=0]{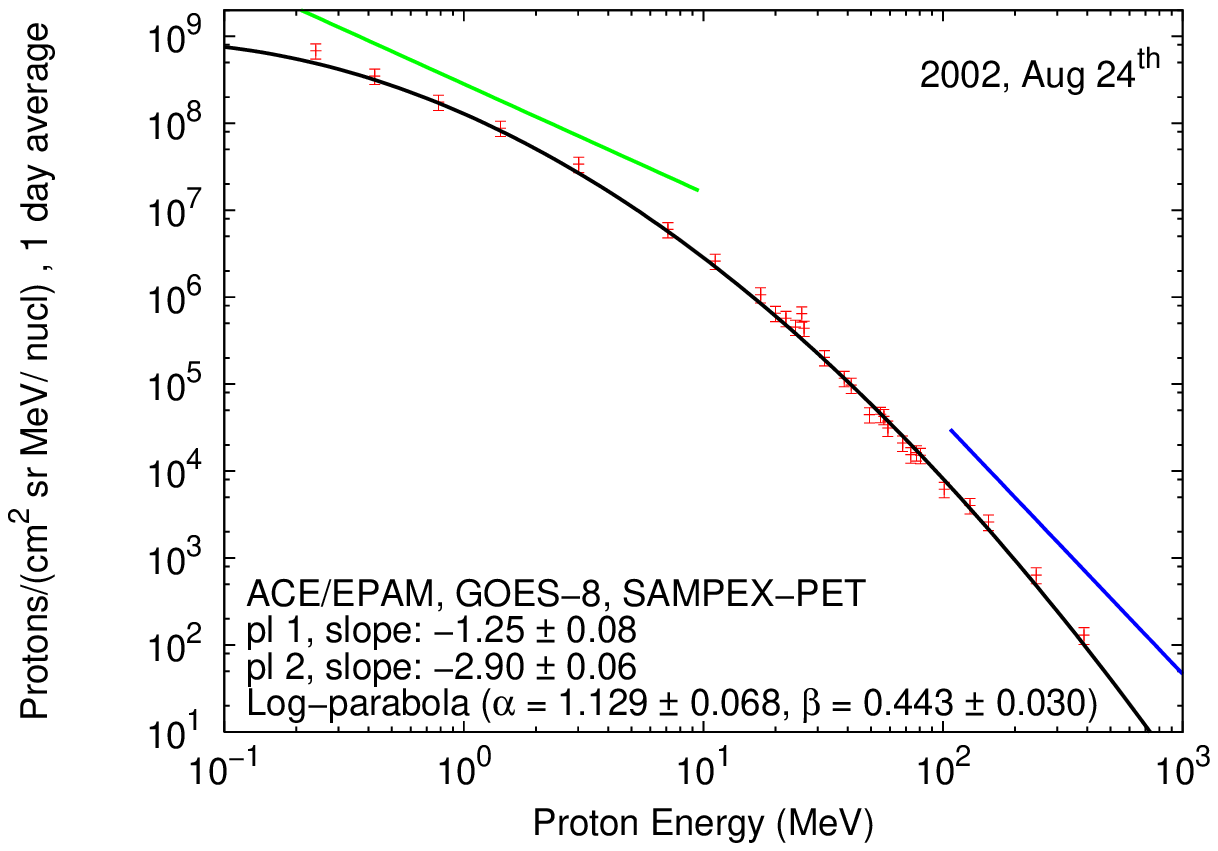}
\includegraphics[scale=0.7,angle=0]{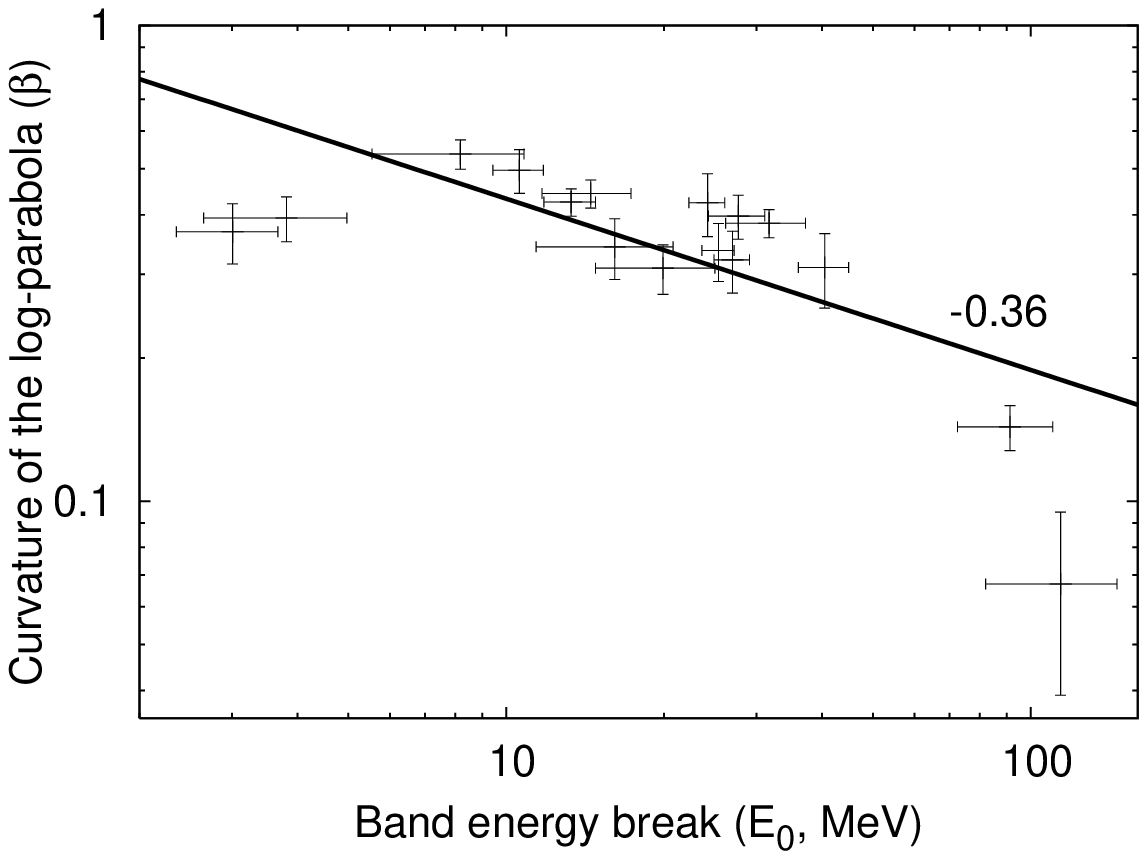}
\caption{{\bf Left:} Fluence $1-$day integrated spectrum as measured by ACE/EPAM, GOES-8 and SAMPEX-PET during the event on August $24^{th}$ 2002 is fitted with a log-parabola spectrum; asymptotic power law indexes from Band function best-fit   \citep{Mewaldt.etal:12} are reported. {\bf Right:} For all 16 GLE events in solar cycle $23$, best-fit curvature of the log-parabola  vs best-fit energy break of the Band function. \label{fig:1}}
\end{center}
\end{figure}

We have used multi-spacecraft $1-$day integrated proton spectra of the GLEs of solar cycle $23$ from ACE/ULEIS, ACE/EPAM, GOES-8/11, SAMPEX-PET instruments, available at the energetic particles repository {http$://$www.srl.caltech.edu$/$sampex$/$
DataCenter$/$DATA$/$EventSpectra$/$index$\_$ace.html}. We have carried out a best-fit of the entire GLE sample. Figure \ref{fig:1}, left panel, shows the log-parabola best-fit of the August $24^{th} $ 2002 event in comparison with the two asymptotic power-laws from the Band function best-fit. Figure \ref{fig:1}, right panel, shows for each of the $16$ events our best-fit $\beta$ vs the best-fit $E_0$ from \citet{Mewaldt.etal:12}. Despite  the limited sample available, we find an anticorrelation of the best-fit parameters. The only event with $E_0 > 100$ MeV has a large relative error for both $\beta$ ($41 \%$) and $E_0$ ($28 \%$) due to the little spectral curvature in the energy range considered, i.e., $0.08  -100 $ MeV. The two events with $E_0 < 5$ MeV have a spectral coverage $\leq 3$ decades with also a resulting poorer determination of $E_0$ (and $\beta$).

In summary, our finding shows for most of the $16$ GLE events of cycle $23$ an anti-correlation $\beta \simeq E_0 ^{-0.36}$. Such an anti-correlation results from the fact that a shock structure confining higher-energy particles (higher $E_0$) enables particle acceleration via the diffusive mechanism up to higher energies with a spectrum closer to a power-law (smaller $\beta$). A log-parabola fit allows also for a reduced number of free parameters, modulo a normalization: one power-law index ($\alpha$) and a global spectral curvature ($\beta$) as opposed to two power-laws indexes and the break $E_0$. Thus, we propose to replace the single energy value of the break $E_0$ with the parameter $\beta$ that describes a global spectral behaviour.

\acknowledgments

We are grateful for the assistance of Dr. R. Mewaldt on multi-spacecraft data and for reading the manuscript. CZ thanks Harvard University for the hospitality and Winsor School ILE program. FF acknowledges support from NASA under Grant NNX15AJ71G and Scholarly Studies Award 40488100HH00181 at Harvard/Smithsonian Center for Astrophysics.


\def \apss{{\it Astrophys.\ Sp.\ Sci.}}
\def \aj{{\it AJ}}
\def \apj{{\it ApJ}}
\def \apjl{{\it ApJL}}
\def \apjs{{\it ApJS}}
\def \araa{{\it Ann. Rev. A \& A}}
\def \prc{{\it Phys.\ Rev.\ C}}
\def \aap{{\it A\&A}}
\def \aaps{{\it A\&ASS}}
\def \mnras{{\it MNRAS}}
\def \physscr{{\it Phys.\ Scripta}}
\def \pasp{{\it Publ.\ Astron.\ Soc.\ Pac.}}
\def \gca{{\it Geochim. Cosmochim.\ Act.}}
\def \nat{{\it Nature}}
\def \solphys{{\it Sol.\ Phys.}}

\bibliographystyle{aasjournal}
\bibliography{rnaas}






\end{document}